\documentclass[pre,twocolumn]{revtex4-1}
\usepackage{amsmath}
\usepackage{graphicx}
\usepackage{color}

\begin{document}

\title{Self-consistent theory for sound propagation in a simple model of a disordered, harmonic solid}
\author{Grzegorz Szamel}
\affiliation{Department of Chemistry,
  Colorado State University, Fort Collins, CO 80523}

\begin{abstract}
  We present a self-consistent theory for sound propagation in a simple model of a disordered solid.
  The solid is modeled as a collection of randomly distributed particles connected by harmonic
  springs with strengths that depend on the interparticle distances, \textit{i.e.} the Euclidean random
  matrix model of M\'{e}zard
  \textit{et al.} [Nucl. Phys. \textbf{559B}, 689 (1999)]. The derivation of the theory combines
  two exact projection operator steps and a factorization approximation. Within our approach the
  square of the speed of sound is non-negative. 
  We expect that an unjamming transition would manifest itself through
  vanishing of the speed of sound. 
\end{abstract}

\maketitle

\section{Introduction}\label{intro}

In the last two decades the physics of sound propagation in amorphous solids has attracted a lot of experimental
\cite{MasciovecchioPRL2006,RufflePRL2008,BaldiPRL2010,BaldiPRL2014},
simulational \cite{GelinNatMat2016,Mizuno2018,Wang2019,Moriel2019}
and theoretical \cite{Buchenau1992,Schober2011,SchirmacherJNCS2011,DeGiuliSM2014} interest.
The inherent disorder of these solids leads to microscopic-scale non-affine response to external stresses
\cite{WittmerEPL2002,TanguyPRB2002,MaloneyPRL,LemaitreJSP}.
These non-affine effects cause a renormalization of elastic constants \cite{LeonfortePRB,LemaitreJSP}
and speeds of sound of amorphous
solids from their values predicted by the classic Born-Huang expressions \cite{BornHuang}.
The non-affine effects also result in sound attenuation, even in the absence of
anharmonicity and thermal effects  \cite{CaroliJCP2020,BaggioliZaccone,GSEFJCP2022}.
The low-temperature sound attenuation is believed to scale
with the wavevector $k$ or the frequency $\omega$ of the sound wave as $k^{d+1}$ or $\omega^{d+1}$ in
$d$ spatial dimensions, which is referred to as Rayleigh scattering scaling \cite{Wang2019,Moriel2019}. 
The interest in sound attenuation and its scaling with the wavevector comes from the fact that these
properties influence other characteristic features of low-temperature amorphous solids; \textit{inter alia},
they can be connected to the density of states and the boson peak
via the so-called generalized Debye model \cite{Mizuno2018}.

There are two mesoscopic approaches to describe sound propagation in amorphous solids.
The first one, local oscillator or soft-potential model \cite{Buchenau1992,Schober2011, BuchenauinRamos},
posits the existence of localized low-energy excitations
that interact with plane waves and lead to sound attenuation.
The local oscillator model has provided a valuable way to rationalize
various experimental results, but it lacks a fundamental microscopic derivation. In particular, the microscopic
interpretation of the local excitations and their identification in simulated amorphous solids is a subject
of ongoing research \cite{Gartner2016,Wijtmans2017,WangNatComm2019,Scalliet2019,Kapteijns2020,Moriel2020}.

The second mesoscopic approach is known as the fluctuating elasticity
theory \cite{SchirmacherEPL,SchirmacherPRL2007,SchirmacherJNCS2011}.
It assumes that an amorphous
solid can be modeled as a continuous medium with spatially varying elastic constants. The spatial variation
of the elastic constants leads to sound scattering and attenuation. In the limit of the sound wavelength 
much larger than the characteristic spatial scale of the elastic inhomogeneities one gets Rayleigh scattering
of sound waves and $k^{d+1}$ scaling of the sound attenuation coefficient. The physical picture
of the fluctuating elasticity theory is clear and appealing and for this reason it has been used
to help with the interpretation of experimental data \cite{Pogna2019}. Importantly,
it is possible to relate the fluctuating elasticity theory to the microscopic, particle-based models
of amorphous solids \cite{CaroliPRL2019}. 
However, practical applications require quantifying local elastic
heterogeneity, which is not unique \cite{MizunoPRE2013}. 
Quantitative comparisons between fluctuating elasticity predictions and
sound attenuation simulations led to widely different opinions about the accuracy of the theory
\cite{CaroliPRL2019,KapteijnsJCP,MahajanPRL2021}.

Recently, we derived a microscopic theory of sound propagation in low-temperature, harmonic
amorphous elastic solids \cite{GSEFJCP2022}. 
We argued that in the long wavelength limit our theory is exact. We verified
this statement by reproducing very accurately both the speeds of sound and sound attenuation coefficients,
which were measured independently via direct sound attenuation simulations \cite{Wang2019},
for a number of amorphous solids with widely varying stabilities. 

To evaluate speeds of sound and sound attenuation coefficients our theory needs eigenvalues and eigenvectors
of the microscopic Hessian matrix. Therefore, our expressions for the speeds of sound and sound
attenuation coefficients are similar to the well-known Green-Kubo expressions \cite{HansenMcDonald} 
for transport coefficients; 
they are exact but they need microscopic information to produce
explicit numerical results. It would be interesting to develop a theory that could lead to 
speeds of sound and sound attenuation coefficients using only limited information about the solid's
structure, \textit{e.g.} the pair distribution function, even if this theory were only approximate.

Here we make the first step towards this goal. We present an approximate self-consistent theory
for the propagation of plane waves in a very simple model of a harmonic, amorphous solid
known as the Euclidean random matrix (ERM) model \cite{MezardNuclPhys199}.
In this model, the particles are distributed randomly and independently.
They are connected by harmonic springs whose strengths depend on the
interparticle distances. The model assumes ``scalar displacements'' of the particles from
their positions. The goal of the theory is to predict the propagation of plane waves of these scalar displacements. 

There have been several studies of the ERM model. Grigera \textit{et al.} \cite{GrigeraPRL2001}
and Martin-Mayor \textit{et al.} \cite{MartinMayorJCP2001}
developed an approximate self-consistent theory for the resolvent of this model. As shown
by Ganter and Schirmacher \cite{GanterPRB2010},
this early theory leads to incorrect predictions for sound attenuation.
The problem was subsequently re-analyzed in Grigera \textit{et al.} \cite{GrigeraJSM2011}
and it was found that the early theory of Refs. \cite{GrigeraPRL2001,MartinMayorJCP2001}
overlooked a non-trivial cancellation of terms in the diagrammatic expansion for the ERM model resolvent and
that the ERM model predicts Rayleigh scattering scaling of the sound attenuation.

Ganter and Schirmacher \cite{GanterPRB2010,GanterPM2001}
also presented two approximate self-consistent theories for the resolvent.
These theories predict Rayleigh scattering scaling of sound attenuation. However, the
theories were postulated rather than derived and it is not clear how one could improve upon them.
The simpler of these theories was later reviewed and analyzed by Schirmacher \textit{et al.}
\cite{SchirmacherJPA2019}.

Recently Vogel and Fuchs \cite{VogelPRL2023}
proposed and analyzed a different self-consistent theory for the resolvent of
the ERM model. This theory was based on a resummation of a diagrammatic expansion
for the resolvent. 

Here we present yet another self-consistent theory for the propagation of scalar plane waves in the ERM model.
The general formulation of our theory is similar to that of Vogel and Fuchs. In both
cases the self-consistency is enforced at the level of the vertex function. The advantage of our theory compared to
that of Ref. \cite{VogelPRL2023} is that it is formulated in such a way that the square of the speed of sound
is always non-negative. Our theory follows the scheme proposed in Ref. \cite{GSPRE2023}, in which we
re-formulated the standard expression for elastic constants of amorphous materials and showed that
these constants are always non-negative. Our theory relies upon a factorization approximation similar to that used
in the mode-coupling theory of glassy dynamics and the glass transition \cite{Goetzebook}. It can be improved upon
by moving the factorization approximation to a higher level, in the spirit of the generalized mode-coupling
theory \cite{SzamelPRL2003,JanssenFinP}.

The derivation of our theory uses the method of projection operators \cite{Zwanzigbook} but it is quite explicit.
Importantly, static correlations do not enter due to the completely random
arrangement of the particles. The goal of future research is to reformulate the derivation in such a way that it
can account for highly non-trivial and non-equilibrium
local static correlations present in realistic models of amorphous solids. 

Recently, a new, general theory for vibrational properties of low-temperature amorphous solids
was presented by Vogel \textit{et al.} \cite{Vogel2024}.
This theory starts from the microscopic dynamics and develops an
approximate description of the transverse current correlations in the limit of zero temperature.
Similarly to our approach, the theory of Ref. \cite{Vogel2024}
relies on a combination of several projection operator steps and a self-consistent approximation
similar to that used in the mode-coupling theory.
We believe that our theory is the ERM relative of the theory presented in Ref. \cite{Vogel2024}. 
In particular, both approaches are formulated in such a way that they naturally result in
non-negative square speeds of sound. 

The paper is organized as follows. In Sec. \ref{problem} we present the ERM model and express
the velocity auto-correlation function in terms of the resolvent of the Hessian matrix.
In Sec. \ref{selfenergy} we derive a formal expression for the self-energy of the model.
In Sec. \ref{vertex} we re-write the self-energy in terms of a vertex function and we  
derive a formal equation for this function. In Sec. \ref{fact} we formulate the factorization approximation for the
higher order self-energy that enters into the equation for the vertex function. In Sec. VI we discuss the
dispersion relation and the sound attenuation. In the Discussion section
we present the outlook for future work.

\section{Statement of the problem}\label{problem}

In the ERM model of an amorphous solid the positions $\mathbf{R}_i$, $i=1,\ldots,N$,
of the particles are randomly and independently distributed. In other words, the probability
distribution of a given configuration reads
\begin{equation}\label{PN}
  P(\mathbf{R}_1,\ldots,\mathbf{R}_N) = \frac{1}{V^N}.
\end{equation}

The particles are connected by springs, with spring constants $k_{ij}$ that depend on interparticle
distances,
\begin{equation}\label{kij}
  k_{ij} = f(R_{ij}),
\end{equation}
where $R_{ij} = |\mathbf{R}_{ij}|$ and $\mathbf{R}_{ij} = \mathbf{R}_i-\mathbf{R}_j$.
    
We consider the scalar version of the ERM model, in which one investigates the time dependence
of  ``scalar displacements'' $\phi_i(t)$ of the particles from their equilibrium positions
(we will often omit the time argument). Assuming, as usual, that the particles have the same mass, $m=1$, 
the equations of motion for the displacements read
\begin{eqnarray}\label{HEOM}
  \partial_t^2 \phi_i = -\sum_j \mathcal{H}_{ij} \phi_j. 
\end{eqnarray}
In Eq. \eqref{HEOM} $\mathcal{H}$ is the Hessian matrix
\begin{eqnarray}\label{H}
\mathcal{H}_{il} &=& 
\delta_{il} \sum_{j\neq i} f(\mathbf{R}_{ij})
- \left(1-\delta_{il}\right)f(\mathbf{R}_{il})
\nonumber \\ &=&
\delta_{il} \sum_{j} f(\mathbf{R}_{ij})
- f(\mathbf{R}_{il}).
\end{eqnarray}
We note that the form of $\mathcal{H}$ in the second line of Eq. \eqref{H} can only be used if $f$ is regular at the
origin.

It follows from the definition of the Hessian that
\begin{eqnarray}\label{H0modes}
\sum_l \mathcal{H}_{il} = \sum_i  \mathcal{H}_{il} = 0,
\end{eqnarray}
which expresses the fact that no net force is induced by a uniform displacement of
all the particles. It is equivalent to stating that a uniform translation is an
eigenvector of the Hessian matrix corresponding to the zero eigenvalue.

If function $f(R)$ has finite support, it is possible that the Hessian 
has additional zero eigenvectors corresponding to subsets of particles moving
independently. This would correspond to an unjammed solid. In this paper we will
implicitly assume the standard Gaussian form of the spring constant for which we do not
expect an unjamming transition. The latter transition was discussed by Vogel \textit{et al.} \cite{Vogel2024}
within an approach that started from the microscopic dynamics. As we mentioned in Sec. \ref{intro},
the theory of Ref. \cite{Vogel2024} is very much related to ours.

To monitor (scalar) sound propagation we adopt the procedure introduced by Gelin \textit{et al.}
\cite{GelinNatMat2016}. 
We assume that at $t=0$ the displacements vanish but their initial velocities correspond to a plane wave, 
\begin{eqnarray}\label{initial1}
\phi_i(t=0) = 0,
\text{\hskip 1em} \dot{\phi}_i(t=0) = e^{-i\mathbf{k}\cdot\mathbf{R}_i}\text{\hskip 1em} i=1,\ldots,N.
\end{eqnarray}
We note that for our disordered harmonic solid the amplitude of the plane wave does not play any role and
in Eq. \eqref{initial1} it was set to 1.
Following Gelin \textit{et al.} we focus on the single-particle velocity autocorrelation function,
averaged over all possible configurations,
\begin{eqnarray}\label{Cdef}
  C(t) = \overline{\dot{\phi}_1^*(t=0) \dot{\phi}_1(t)}
    \equiv \frac{1}{N} \sum_{i=1}^N \overline{\dot{\phi}_i^*(t=0)\dot{\phi}_i(t)}.
\end{eqnarray}
Here the overline $\overline{ \ldots}$ denotes averaging over probability distribution \eqref{PN} of the positions
of the particles.

The second equality
in \eqref{Cdef} follows from the fact that after sample averaging the system becomes
homogeneous and thus $C(t)$ can be averaged over the whole system (which in practice
improves the statistics). However, in the theoretical development we will use the
original form of $C(t)$, \textit{i.e.} $C(t) = \overline{\dot{\phi}_1^*(t=0) \dot{\phi}_1(t)}$.

To get the renormalized sound velocities and the sound damping one has to investigate
the limit of small wavevectors $\mathbf{k}$. We anticipate that 
we would get damped oscillations 
$C(t)\propto \cos(v k t)\exp(-\Gamma(k)t/2)$, and we identify $v$ as the speed 
of sound and $\Gamma(k)$ as the damping coefficient.

We start by introducing the Fourier transform,
\begin{equation}\label{FTdef}
  f(\omega) = \int_0^\infty f(t) \exp(i(\omega+i\epsilon)t).
\end{equation}

The Fourier transform of autocorrelation function \eqref{Cdef} can be derived
by transforming equations of motion \eqref{HEOM},
\begin{eqnarray}\label{HEOMo1}
  -(\omega+i\epsilon)^2 \phi_i(\omega) =
  -\sum_j \mathcal{H}_{ij} \cdot \phi_j(\omega)\
  + e^{-i\mathbf{k}\cdot\mathbf{R}_i}.
\end{eqnarray}
We formally solve Eq. \eqref{HEOMo1} to get
\begin{eqnarray}\label{HEOMo2}
  \phi_1(\omega) =
  - \sum_j \left[\left(\omega+i\epsilon\right)^2 \mathcal{I} - \mathcal{H}\right]_{1j}^{-1}
    e^{-i\mathbf{k}\cdot\mathbf{R}_j},
\end{eqnarray}
where $\mathcal{I}$ is the unit tensor, $\mathcal{I}_{ij}=\delta_{ij}$,  
and then use this solution to write the formal expression for the Fourier transform of the 
velocity autocorrelation function,
\begin{eqnarray}\label{HEOMo3}
  \dot{\phi}_1(\omega) = i(\omega+i\epsilon)
  \sum_j \left[\left(\omega+i\epsilon\right)^2 \mathcal{I} - \mathcal{H}\right]_{1j}^{-1}
    e^{-i\mathbf{k}\cdot\mathbf{R}_j},
\end{eqnarray}
and
\begin{eqnarray}\label{Codef}
  C(\omega) &=&  i\left(\omega+i\epsilon\right) \overline{ \sum_{j}  
    e^{i\mathbf{k}\cdot\mathbf{R}_1}\cdot
    \left[\left(\omega+i\epsilon\right)^2 \mathcal{I} - \mathcal{H}\right]_{1j}^{-1}
    e^{-i\mathbf{k}\cdot\mathbf{R}_j}
  }
  \nonumber \\ &=&
  \frac{i\left(\omega+i\epsilon\right)}{N} \overline{ \sum_{i,j}  
    e^{i\mathbf{k}\cdot\mathbf{R}_i}\cdot
    \left[\left(\omega+i\epsilon\right)^2 \mathcal{I} - \mathcal{H}\right]_{ij}^{-1}
    e^{-i\mathbf{k}\cdot\mathbf{R}_j}
  }.
  \nonumber \\
\end{eqnarray}
We note that the Fourier transform \eqref{Codef} 
is related to the resolvent of the Hessian matrix, which is defined as 
\begin{eqnarray}\label{resdef}
G(\mathbf{k};z) =  \frac{1}{N} \overline{
    \sum_{i,j} e^{i\mathbf{k}\cdot\mathbf{R}_i}
    \left[z\mathcal{I}-\mathcal{H}\right]_{ij}^{-1}
    e^{-i\mathbf{k}\cdot\mathbf{R}_j}
}.
\end{eqnarray}
Specifically, $C(\omega)$ can be expressed in terms of the resolvent at $z=\left(\omega+i\epsilon\right)^2$,
\begin{equation}\label{Cores}
  C(\omega) =  i\left(\omega+i\epsilon\right)
  G\left(\mathbf{k};\left(\omega+i\epsilon\right)^2\right).
\end{equation}
Thus, in order to analyze sound propagation we need to develop a theory for the resolvent. 
In the next two sections we will first express the resolvent in terms of a new self-energy function,
then we will express this self-energy in terms of a vertex function and finally we will analyze the latter function.

\section{Self-energy}\label{selfenergy}

We note that expression \eqref{resdef} for the resolvent resembles formal expressions
encountered in theories for stochastic dynamics of interacting objects, \textit{e.g.} dynamics of colloidal
particles \cite{CHess,SzamelLoewen} or Glauber dynamics of interacting spins \cite{Kawasaki}.
Readers familiar with these
theories will undoubtely notice the similarity between the transformations used in this and the
next sections and derivations of theories for colloidal dynamics. However, no knowledge of the
latter subject is necessary for the understanding the present paper.

We start with the formula for the resolvent without explicit averaging over different sites,
\begin{eqnarray}\label{resdef2}
G(\mathbf{k};z) =   \overline{ \sum_{j} 
   e^{i\mathbf{k}\cdot\mathbf{R}_1}
    \left[z\mathcal{I}-\mathcal{H}\right]_{1j}^{-1}
    e^{-i\mathbf{k}\cdot\mathbf{R}_j}
  }.
\end{eqnarray}

To make the notation more compact we define the Fourier transform of the density associated
with site $i$,
\begin{eqnarray}\label{nidef}
  n_i(\mathbf{k}) = e^{-i\mathbf{k}\cdot\mathbf{R}_i}.
\end{eqnarray}
Next, we define projection operator $\mathcal{P}_i$ on the density associated with site $i$,
\begin{eqnarray}\label{Pidef}
  \mathcal{P}_i f =  n_i(\mathbf{k}) \overline{n_i(-\mathbf{k}) f}
  \equiv \frac{n_i(\mathbf{k})}{N}\sum_j \overline{n_j(-\mathbf{k}) f},
\end{eqnarray}
where we used the fact that the averaged expression,  $\overline{e^{i\mathbf{k}\cdot\mathbf{R}_i}f}$,
is site-independent. We also define orthogonal projection, $\mathcal{Q}_i$, 
\begin{eqnarray}\label{Qidef}
  \mathcal{Q}_if = f- n_i(\mathbf{k}) \overline{n_i(-\mathbf{k}) f}.
\end{eqnarray}
To make some future equations more compact it is convenient to introduce an alternative symbol for
averaging over the probability distribution of the sites,
\begin{eqnarray}\label{Pdef}
  \mathcal{P} f = \overline{f}.
\end{eqnarray}

We start the analysis of the resolvent by writing equation for
$zG-1$,
\begin{eqnarray}\label{res1}
  zG(\mathbf{k};z) - 1 =
    \overline{ \sum_{j,k} n_1(-\mathbf{k})  \mathcal{H}_{1j}
    \left[z\mathcal{I}-\mathcal{H}\right]_{jk}^{-1} n_k(\mathbf{k}) }.
\end{eqnarray}

Next, we use a series of projection operator steps detailed in Appendix \ref{appA} and we arrive at the following
expression for $zG-1$,
\begin{widetext}
\begin{eqnarray}\label{res8}
  &&  zG(\mathbf{k};z) - 1 =  \rho \left(f(0)-f(k)\right) G(\mathbf{k};z)
  \nonumber \\ &&
  +  \overline{ \sum_{j}  
    n_1(-\mathbf{k})  \mathcal{H}_{1j} \mathcal{Q}_j 
  \sum_l \left\{z\mathcal{I} - \mathcal{Q} \mathcal{H}\mathcal{Q}
  +\mathcal{Q} \mathcal{H} n  
  \left[\rho \left(f(0)-f(k)\right) \right]^{-1}
  \mathcal{P} n \mathcal{H}\mathcal{Q}
  \right\}^{-1}_{jl} 
  \sum_{m} \mathcal{Q}_l \mathcal{H}_{lm} 
  n_m(\mathbf{k})  }
  \nonumber \\ &&
  \times \left[\rho \left(f(0)-f(k)\right) \right]^{-1} \left[zG(\mathbf{k};z) - 1\right]
\end{eqnarray}
\end{widetext}

To make the subsequent equations more compact we define the projected Hessian, which plays the role of 
the evolution operator in Eq. \eqref{res8},
\begin{eqnarray}\label{Hirr1}
  && \mathcal{H}^\Sigma_{ij} =
  \mathcal{Q}_i \mathcal{H}_{ij}\mathcal{Q}_j
  \\ \nonumber &&
  - \sum_k \mathcal{Q}_i \mathcal{H}_{ik} n_k(\mathbf{k})
  \left[\rho \left(f(0)-f(k)\right) \right]^{-1}
  \mathcal{P} n_k(-\mathbf{k})\mathcal{H}_{kj}\mathcal{Q}_j.
\end{eqnarray}
The superscript $\Sigma$ anticipates the fact that the projected Hessian 
determines time evolution of the self-energy $\Sigma$.

We define the self-energy as follows
\begin{eqnarray}\label{seirr1}
  && \!\! \Sigma(\mathbf{k};z) \! = \!
  - \overline{ \sum_{j,l,m} 
    n_1(-\mathbf{k})  \mathcal{H}_{1j}
    \mathcal{Q}_j  \left[ z\mathcal{I} - \mathcal{H}^\Sigma \right]^{-1}_{jl} 
  \mathcal{Q}_l\mathcal{H}_{lm}n_m(\mathbf{k})}
  \nonumber \\ &&  \!\! \equiv 
  - \frac{1}{N} \overline{ \sum_{i,j,l,m} 
    n_i(-\mathbf{k})  \mathcal{H}_{ij}
    \mathcal{Q}_j  \left[ z\mathcal{I} - \mathcal{H}^\Sigma \right]^{-1}_{jl} 
  \mathcal{Q}_l\mathcal{H}_{lm}n_m(\mathbf{k})}
  \nonumber \\
\end{eqnarray}

Using the self-energy, we can express the resolvent in the following way,
\begin{eqnarray}\label{res12}
  G(\mathbf{k};z) =
  \frac{1}{z- \frac{\rho \left(f(0)-f(k)\right)}
    {1+ \Sigma(\mathbf{k};z) \left[\rho\left(f(0)-f(k)\right)\right]^{-1}}}.
\end{eqnarray}

We finish this section by noting that our definition of the self-energy, Eq. \eqref{seirr1}, and
our expression of the resolvent in terms of the self-energy, Eq. \eqref{res12}, differ
from those used by previous researchers.
The self-energy we defined is the analogue of the \textit{irreducible} memory function
that was introduced in the context of colloidal dynamics by Cichocki and Hess \cite{CHess}
and then discussed in a more general setting by Kawasaki \cite{Kawasaki}. 
For this reason, we could have named the quantity defined in Eq. \eqref{seirr1} the
irreducible self-energy. However, since the name self-energy already implies irreducibility,
we decided not to add the adjective ``irreducible''.

Following the analysis of Ganter and Schirmacher \cite{GanterPRB2010}
and of Vogel and Fuchs \cite{VogelPRL2023},
we anticipate that in order to predict Rayleigh scattering scaling of the sound
attenuation we cannot use a factorization approximation in expression
\eqref{seirr1}. Instead, in the next section we will rewrite the self-energy in terms of
a vertex function and we will express this function in terms of its own order self-energy.
Finally, in Sec. \ref{fact} we will apply a factorization approximation to the self-energy
associated with the vertex function.

\section{Vertex function}\label{vertex}

First, the expression at the
right-hand-side of the definition of the self-energy, Eq. \eqref{seirr1}, can 
be re-written as follows,
\begin{eqnarray}\label{ver2}
  && \sum_m \mathcal{Q}_l\mathcal{H}_{lm}n_m(\mathbf{k})
  \equiv \sum_m \mathcal{Q}_l\mathcal{H}_{lm} e^{-i\mathbf{k}\cdot\mathbf{R}_m} 
  \\ \nonumber &=&
  \frac{1}{V}\sum_{\mathbf{q}_2}\left( f(\mathbf{k}-\mathbf{q}_2) - f(q_2) \right)
  \mathcal{Q}_l \sum_m e^{-i(\mathbf{k}-\mathbf{q}_2)\cdot\mathbf{R}_{ml}} e^{-i\mathbf{k}\cdot\mathbf{R}_l}
\end{eqnarray}
We note that the expression in the second  line above involves the part of the product of the density of site $l$,
$e^{-i\mathbf{q}_2\cdot\mathbf{R}_l}=n_l(\mathbf{q}_2)$, 
and the collective density of all sites,
$\sum_m e^{-i(\mathbf{k}-\mathbf{q}_2)\cdot\mathbf{R}_m}=n(\mathbf{k}-\mathbf{q}_2)$, that is orthogonal
to the density of site $l$. Using definition \eqref{Qidef} of orthogonal projection $\mathcal{Q}_l$
one can show that this quantity is equal to the product of the density of site $l$ and
the fluctuation of the collective density of all sites \textit{different} from $l$.  We will denote this product by 
$n_{l2}$,
\begin{eqnarray}\label{denprod}
  && \mathcal{Q}_l \sum_m e^{-i\left(\mathbf{k}-\mathbf{q}_2\right)\cdot\mathbf{R}_{ml}} e^{-i\mathbf{k}\cdot\mathbf{R}_l}
  \nonumber \\ &=&
  \mathcal{Q}_l e^{-i\mathbf{q}_2\cdot\mathbf{R}_l} \sum_m e^{-i(\mathbf{k}-\mathbf{q}_2)\cdot\mathbf{R}_m}
  \nonumber \\ &=&
  e^{-i\mathbf{q}_2\cdot\mathbf{R}_l} \left[\sum_{m\neq l} e^{-i\left(\mathbf{k}-\mathbf{q}_2\right)\cdot\mathbf{R}_m} -
    \overline{\sum_{m\neq l} e^{-i\left(\mathbf{k}-\mathbf{q}_2\right)\cdot\mathbf{R}_m}}\right] 
 \nonumber \\ &\equiv & n_{l2}(\mathbf{q}_2,\mathbf{k}-\mathbf{q}_2).
\end{eqnarray}
Similarly, one can show that the expression at the left-hand-side of Eq. \eqref{seirr1} can be
re-written as
\begin{eqnarray}\label{ver3}
  && \sum_i n_i(-\mathbf{k}) \mathcal{H}_{ij} \mathcal{Q}_j
  \\ \nonumber &=&
  \frac{1}{V}\sum_{\mathbf{q}_1}\left( f(\mathbf{k}-\mathbf{q}_1) - f(q_1) \right)
  n_{j2}(-\mathbf{q}_1,\mathbf{q}_1-\mathbf{k}).
\end{eqnarray}
  
Equations \eqref{ver2} and \eqref{ver3} allow us to express the self-energy in term of the vertex function,
\begin{eqnarray}\label{seirr2}
  && \Sigma(\mathbf{k};z) =
  \frac{1}{V} \sum_{\mathbf{q}_1}
  \left( f(\mathbf{k}-\mathbf{q}_1)-f(q_1) \right)
  \mathcal{V}(\mathbf{q}_1,\mathbf{k};z)
\end{eqnarray}
where vertex function $\mathcal{V}(\mathbf{q}_1,\mathbf{k};z)$ reads
\begin{eqnarray}\label{renver1}
  && \!\!\! \mathcal{V}(\mathbf{q}_1,\mathbf{k};z) =
  - \frac{1}{V} 
  \sum_{\mathbf{q}_2}
  \nonumber \\ && \!\!\! \times
  \overline{\sum_{l} n_{j2}(-\mathbf{q}_1,\mathbf{q}_1-\mathbf{k})
    \left[z\mathcal{I} - \mathcal{H}^\Sigma \right]^{-1}_{jl}
    n_{l2}(\mathbf{q}_2,\mathbf{k}-\mathbf{q}_2)}
  \nonumber \\ && \!\!\! \times
  \left( f(\mathbf{k}-\mathbf{q}_2) - f(q_2) \right).
\end{eqnarray}

The analysis of the vertex function parallels the analysis of the resolvent. 
We will first define a new projection operator that projects on $n_{j2}$,
\textit{i.e.} on the product of the single site density of
site $j$ and the fluctuation of the collective density of all other sites. Then, we will write down an equation for
$z\mathcal{V}(\mathbf{q}_1,\mathbf{k};z)-\lim_{z\to\infty}z\mathcal{V}(\mathbf{q}_1,\mathbf{k};z)$
and insert the identity written as the sum of the projection and the orthogonal projection
into this equation. Next, we will analyze the resulting equation and define a self-energy for the vertex function.
Finally, we will express the vertex function in terms of its self-energy in an equation analogous
to Eq. \eqref{res12}.

The second projection operator is the projection on $n_{j2}$, which is the product of the single site density of site
$j$ and the fluctuation of the collective density of all other sites,
\begin{eqnarray}\label{Pj2def}
  \mathcal{P}_{j2}^{\mathbf{q}_1} f &=& \sum_{\mathbf{q}_2}
  n_{j2}(\mathbf{q}_1,\mathbf{k}-\mathbf{q}_1)
  \nonumber \\ && \times \overline{n_{j2}(-\mathbf{q}_1,-\mathbf{k}+\mathbf{q}_1)
    n_{j2}(\mathbf{q}_2,\mathbf{k}-\mathbf{q}_2)}^{-1}
  \nonumber \\ && \times  \overline{ n_{j2}(-\mathbf{q}_2,-\mathbf{k}+\mathbf{q}_2) f}.
\end{eqnarray}
In Appendix \ref{appB} we show that the absence of any static correlations allows us to re-write
Eq. \eqref{Pj2def} as follows,
\begin{eqnarray}\label{Pj2def2}
  \mathcal{P}_{j2}^{\mathbf{q}_1} f
  =
  \frac{1}{N}n_{j2}(\mathbf{q}_1,\mathbf{k}-\mathbf{q}_1)
  \overline{ n_{j2}(-\mathbf{q}_1,-\mathbf{k}+\mathbf{q}_1) f}.
  \nonumber \\
\end{eqnarray}
We also define the projection on the space orthogonal to the space spanned by functions $n_{j2}$,
\begin{eqnarray}\label{Qj2def2}
  \mathcal{Q}_{j2} f
  = 
  f - \frac{1}{N} \sum_{\mathbf{q}_1}
   n_{j2}(\mathbf{q}_1,\mathbf{k}-\mathbf{q}_1)
   \overline{ n_{j2}(-\mathbf{q}_1,-\mathbf{k}+\mathbf{q}_1) f}.
   \nonumber \\
\end{eqnarray}

To simplify the notation, we define auxiliary quantity $\mathcal{V}_0$,
\begin{eqnarray}\label{ver0def}
  \mathcal{V}_0(\mathbf{q}_1,\mathbf{k}) = \lim_{z\to\infty} z \mathcal{V}(\mathbf{q}_1,\mathbf{k};z).
\end{eqnarray}
In Appendix \ref{appC} we show that 
\begin{eqnarray}\label{ver0res}
  \mathcal{V}_0(\mathbf{q}_1,\mathbf{k}) = - \rho \left(f(\mathbf{k}-\mathbf{q}_1) - f(q_1) \right).
\end{eqnarray}

The analysis of the vertex function begins with writing
equation for $z\mathcal{V}-\mathcal{V}_0$ (note the analogy with Eq. \eqref{res1}),
\begin{widetext}
\begin{eqnarray}\label{renver3}
  z\mathcal{V}(\mathbf{q}_1,\mathbf{k};z) - \mathcal{V}_0(\mathbf{q}_1,\mathbf{k}) =
  - \frac{1}{V} 
  \sum_{\mathbf{q}_2}
  \overline{\sum_{n,l} n_{j2}(-\mathbf{q}_1,-\mathbf{k}+\mathbf{q}_1)
    \mathcal{H}^\Sigma_{jn}
    \left[z\mathcal{I} - \mathcal{H}^\Sigma \right]^{-1}_{nl} 
    n_{l2}(\mathbf{q}_2,\mathbf{k}-\mathbf{q}_2)}
  \left( f(\mathbf{k}-\mathbf{q}_2) - f(-\mathbf{q}_2) \right).
  \nonumber \\
\end{eqnarray}

The right-hand-side of Eq. \eqref{renver3} is analyzed using projection operators
$\mathcal{P}_{j2}^{\mathbf{q}_1}$ and $\mathcal{Q}_{j2}$. The result of this analysis,
which is presented in Appendix \ref{appD}, is the following expression for
$z\mathcal{V}-\mathcal{V}_0$ (note the analogy with Eq. \eqref{res8}),
\begin{eqnarray}\label{renver8}
  && z\mathcal{V}(\mathbf{q}_1,\mathbf{k};z) - \mathcal{V}_0(\mathbf{q}_1,\mathbf{k}) =
   \sum_{\mathbf{q}_3}\mathcal{F}_{\mathbf{k}}(\mathbf{q}_1,\mathbf{q}_3)
   \mathcal{V}(\mathbf{q}_3,\mathbf{k};z) 
  \nonumber \\ &&
  +\frac{1}{N} 
  \sum_{\mathbf{q}_2}
  \overline{\sum_{n,l} n_{j2}(-\mathbf{q}_1,-\mathbf{k}+\mathbf{q}_1)
    \mathcal{H}^\Sigma_{jn}\mathcal{Q}_{n2}
    \left[z\mathcal{I}-\mathcal{H}^{\mathcal{V}}\right]_{nl}^{-1} \sum_m
    \mathcal{Q}_{l2}\mathcal{H}^\Sigma_{lm} 
    n_{m2}(\mathbf{q}_2,\mathbf{k}-\mathbf{q}_2) }
  \nonumber \\ && \times
  \sum_{\mathbf{q}_3} \left[\mathcal{F}_\mathbf{k}(\mathbf{q}_2,\mathbf{q}_3) \right]^{-1}
  \left[z\mathcal{V}(\mathbf{q}_3,\mathbf{k};z) - \mathcal{V}_0(\mathbf{q}_3,\mathbf{k})\right].
\end{eqnarray}

In Eq. \eqref{renver8} $\mathcal{H}^{\mathcal{V}}$ is the second
projected Hessian (note the analogy with Eq. \eqref{Hirr1}),
\begin{eqnarray}\label{Hirr2main}
  \mathcal{H}^{\mathcal{V}}_{ij} =
  \mathcal{Q}_{i2} \mathcal{H}^\Sigma_{ij} \mathcal{Q}_{j2} -
  \frac{1}{N} \sum_{\mathbf{q}_1,\mathbf{q}_3} \sum_{k} \mathcal{Q}_{i2} \mathcal{H}^\Sigma_{ik}  
  n_{k2}(\mathbf{q}_1,\mathbf{k}-\mathbf{q}_1)
  \left[\mathcal{F}_\mathbf{k}(\mathbf{q}_1,\mathbf{q}_3) \right]^{-1}
  \mathcal{P}
  n_{k2}(-\mathbf{q}_3,-\mathbf{k}+\mathbf{q}_3)
  \mathcal{H}^\Sigma_{kj}
  \mathcal{Q}_{j2}.
\end{eqnarray}

We define the self-energy for the vertex function as follows
\begin{eqnarray}\label{seirr3}
  \Sigma^{\mathcal{V}}(\mathbf{q}_1,\mathbf{q}_2,\mathbf{k};z) &=& 
  - \frac{1}{N} 
  \overline{ \sum_{n,l,m} n_{j2}(-\mathbf{q}_1,-\mathbf{k}+\mathbf{q}_1)
    \mathcal{H}^\Sigma_{jn}\mathcal{Q}_{n2}
    \left[z\mathcal{I}-\mathcal{H}^{\mathcal{V}}\right]_{nl}^{-1} 
    \mathcal{Q}_{l2}\mathcal{H}^\Sigma_{lm} 
    n_{m2}(\mathbf{q}_2,\mathbf{k}-\mathbf{q}_2) }
  \nonumber \\ &\equiv &
  - \frac{1}{N^2} 
  \overline{ \sum_{j,n,l,m} n_{j2}(-\mathbf{q}_1,-\mathbf{k}+\mathbf{q}_1)
    \mathcal{H}^\Sigma_{jn}\mathcal{Q}_{n2}
    \left[z\mathcal{I}-\mathcal{H}^{\mathcal{V}}\right]_{nl}^{-1} 
    \mathcal{Q}_{l2}\mathcal{H}^\Sigma_{lm} 
    n_{m2}(\mathbf{q}_2,\mathbf{k}-\mathbf{q}_2)},
\end{eqnarray}
and then we re-write Eq. \eqref{renver8} 
for the vertex function in terms of its own self-energy,
\begin{eqnarray}\label{renver9}
  z\mathcal{V}(\mathbf{q}_1,\mathbf{k};z) - \mathcal{V}_0(\mathbf{q}_1,\mathbf{k}) =
   \sum_{\mathbf{q}_3}\mathcal{F}_{\mathbf{k}}(\mathbf{q}_1,\mathbf{q}_3)
   \mathcal{V}(\mathbf{q}_3,\mathbf{k};z) 
  -\sum_{\mathbf{q}_2} \Sigma^{\mathcal{V}}(\mathbf{q}_1,\mathbf{q}_2,\mathbf{k};z)
  \sum_{\mathbf{q}_3} \left[\mathcal{F}_\mathbf{k}(\mathbf{q}_2,\mathbf{q}_3) \right]^{-1}
  \left[z\mathcal{V}(\mathbf{q}_3,\mathbf{k};z) - \mathcal{V}_0(\mathbf{q}_3,\mathbf{k})\right].
  \nonumber \\
\end{eqnarray}
\end{widetext}

Eq. \eqref{renver9} corresponds to Eq. \eqref{res12} at the previous level of the projection operator
method. It looks considerably more complicated since it describes a quantity that depends on
two wavevectors, vertex function $\mathcal{V}(\mathbf{q}_1,\mathbf{k};z)$. Even for the very
simple ERM model with completely random arrangement of particles, this fact leads to the appearance
of kernel $\mathcal{F}_\mathbf{k}(\mathbf{q}_1,\mathbf{q}_2)$ and self-energy
$\Sigma^{\mathcal{V}}(\mathbf{q}_1,\mathbf{q}_2,\mathbf{k};z)$ that both 
depend on three wavevectors.

We note that Eq. \eqref{renver9} is similar to the equation derived at the last step of the
projection operator procedure of Vogel \textit{et al.} \cite{Vogel2024}. The main difference is the fact that the
authors of Ref. \cite{Vogel2024} used an approximation that made the analog of kernel
$\mathcal{F}_{\mathbf{k}}(\mathbf{q}_1,\mathbf{q}_2)$ diagonal, \textit{i.e.} proportional
to $\delta_{\mathbf{q}_1,\mathbf{q}_2}$. 

As established by Ganter and Schirmacher \cite{GanterPRB2010}
and by Vogel and Fuchs \cite{VogelPRL2023}, a factorization approximation
at the  previous level leads to a qualitatively incorrect result for the dependence of the sound
attenuation on the wavevector. For this reason, in the next section we will formulate a factorization
approximation for self-energy $\Sigma^{\mathcal{V}}(\mathbf{q}_1,\mathbf{q}_2,\mathbf{k};z)$.
This factorization approximation will result in a system of self-consistent equations similar to
that obtained diagrammatically by Vogel and Fuchs \cite{VogelPRL2023}. The important
difference is that the authors of Ref. \cite{VogelPRL2023} used different definitions of
self-energy for both the resolvent and the vertex function. Our definitions are analogous to
those adopted by Vogel \textit{et al.} in Ref. \cite{Vogel2024}.

\section{Factorization approximation}\label{fact}

To formulate a factorization approximation we first rewrite the formal expression for the
self-energy for the vertex function in terms of many-particle densities.

It is convenient to start with the expression at the right-hand-side of the definition of the definition
self-energy for the vertex function, Eq. \eqref{seirr3},
\begin{eqnarray}\label{vertex1}
  && \sum_{m}  
  \mathcal{Q}_{l2}\mathcal{H}^\Sigma_{lm} n_{m2}(\mathbf{q}_2,\mathbf{k}-\mathbf{q}_2)
  \nonumber \\ &=&
  \sum_m \mathcal{Q}_{l2}\mathcal{Q}_l \mathcal{H}_{lm}
  n_{m2}(\mathbf{q}_2,\mathbf{k}-\mathbf{q}_2)
  \nonumber \\ &&
  - \sum_{k,m} \mathcal{Q}_{l2}\mathcal{Q}_l \mathcal{H}_{lk} n_k(\mathbf{k})
  \left[\rho \left(f(0)-f(k)\right) \right]^{-1}
  \nonumber \\ && \times
  \mathcal{P} n_k(-\mathbf{k})\mathcal{H}_{km} n_{m2}(\mathbf{q}_2,\mathbf{k}-\mathbf{q}_2).
\end{eqnarray}

First, we note that due to the fact that $\mathcal{Q}_{l2}$ projects on the space orthogonal
to that spanned by $n_{l2}$, the second term at the right-hand-side of Eq. \eqref{vertex1} vanishes.
Thus, the non-vanishing contribution will arise from 
\begin{eqnarray}\label{vertex2}
  \sum_m \mathcal{Q}_{l2}\mathcal{Q}_l \mathcal{H}_{lm}
  n_{m2}(\mathbf{q}_2,\mathbf{k}-\mathbf{q}_2).
\end{eqnarray}
In Appendix \ref{appE} we show that the above expression can be expressed in terms of a product
of a single-particle density and two fluctuations of collective densities of other particles,
\begin{eqnarray}\label{nl3}
  && n_{l3}(\mathbf{q}_4,\mathbf{q}_2-\mathbf{q}_4,\mathbf{k}-\mathbf{q}_2) =
  e^{-i\mathbf{q}_4\cdot\mathbf{R}_l}
  \nonumber \\ && \times
  \sum_{n\neq l} \sum_{k\neq l,k\neq n}
  \left[e^{-i(\mathbf{q}_2-\mathbf{q}_4)\cdot\mathbf{R}_{n}} -
    \overline{e^{-i(\mathbf{q}_2-\mathbf{q}_4)\cdot\mathbf{R}_{n}}}\right]
  \nonumber \\ && \times
  \left[e^{-i\left(\mathbf{k}-\mathbf{q}_2\right)\cdot\mathbf{R}_k} -
    \overline{e^{-i\left(\mathbf{k}-\mathbf{q}_2\right)\cdot\mathbf{R}_k}}\right].
\end{eqnarray}
In terms of $n_{l3}$ expression \eqref{vertex2} reads
\begin{eqnarray}\label{vertex3}
  && \frac{1}{V} \sum_{\mathbf{q}_4} \left(f(\mathbf{q}_2-\mathbf{q}_4) - f(q_4) \right)
  n_{l3}(\mathbf{q}_4,\mathbf{q}_2-\mathbf{q}_4,\mathbf{k}-\mathbf{q}_2).
  \nonumber  \\
\end{eqnarray}

Similarly, we can show that the expression
at the left-hand-side of definition \eqref{seirr3} can be rewritten as follows,
\begin{eqnarray}\label{vertex4}
  && \sum_{j} 
  n_{j2}(-\mathbf{q}_1,-\mathbf{k}+\mathbf{q}_1)
    \mathcal{Q}_j \mathcal{H}^\Sigma_{jn}\mathcal{Q}_{n2}
    \\ \nonumber  &=&
    \frac{1}{V} \sum_{\mathbf{q}_3} \left(f(\mathbf{q}_1-\mathbf{q}_3) - f(q_3) \right)
    \\ \nonumber  && \times
    n_{n3}(-\mathbf{q}_3,-\mathbf{q}_1+\mathbf{q}_3,-\mathbf{k}-\mathbf{q}_1).
\end{eqnarray}

Substituting expressions \eqref{vertex3} and \eqref{vertex4} into Eq. \eqref{seirr3} we get
\begin{widetext}
\begin{eqnarray}\label{seirr4}
  \Sigma^{\mathcal{V}}(\mathbf{q}_1,\mathbf{q}_2,\mathbf{k};z) &=& 
  - \frac{1}{N^2 V^2} 
  \sum_n 
  \sum_l 
  \sum_{\mathbf{q}_3} \left(f(\mathbf{q}_1-\mathbf{q}_3) - f(q_3) \right)
  \sum_{\mathbf{q}_4} \left(f(\mathbf{q}_2-\mathbf{q}_4) - f(q_4) \right)
  \nonumber \\ && \times
  \overline{n_{n3}(-\mathbf{q}_3,-\mathbf{q}_1+\mathbf{q}_3,-\mathbf{k}-\mathbf{q}_1)
    \left[z\mathcal{I}-\mathcal{H}^{\mathcal{V}}\right]_{nl}^{-1}
    n_{l3}(\mathbf{q}_4,\mathbf{q}_2-\mathbf{q}_4,\mathbf{k}-\mathbf{q}_2).
    }
\end{eqnarray}
At this point we apply a mode-coupling-like factorization approximation. Specifically, we factorize \emph{and}
we replace projected Hessian $\mathcal{H}^{\mathcal{V}}$ by the Hessian. This approximation allows us
to express the self-energy for the vertex function in terms of the resolvent,
\begin{eqnarray}\label{seirr5}
  \Sigma^{\mathcal{V}}(\mathbf{q}_1,\mathbf{q}_2,\mathbf{k};z) &=& 
  - \frac{1}{N^2 V^2} 
  \sum_n \sum_l 
  \sum_{\mathbf{q}_3} \left(f(\mathbf{q}_1-\mathbf{q}_3) - f(q_3) \right)
  \sum_{\mathbf{q}_4} \left(f(\mathbf{q}_2-\mathbf{q}_4) - f(q_4) \right)
  \nonumber \\ && \times
  \overline{n_{n3}(-\mathbf{q}_3,-\mathbf{q}_1+\mathbf{q}_3,-\mathbf{k}-\mathbf{q}_1)
    \left[z\mathcal{I}-\mathcal{H}^{\mathcal{V}}\right]_{nl}^{-1}
    n_{l3}(\mathbf{q}_4,\mathbf{q}_2-\mathbf{q}_4,\mathbf{k}-\mathbf{q}_2)
    }
  \nonumber \\ &\approx &
  - \frac{1}{N^2 V^2} 
  \sum_n \sum_l 
  \sum_{\mathbf{q}_3} \left(f(\mathbf{q}_1-\mathbf{q}_3) - f(q_3) \right)
  \sum_{\mathbf{q}_4} \left(f(\mathbf{q}_2-\mathbf{q}_4) - f(q_4) \right)
  \nonumber \\ && \times
  \overline{
    n_2(-\mathbf{q}_1+\mathbf{q}_3,-\mathbf{k}-\mathbf{q}_1)
    n_2(\mathbf{q}_2-\mathbf{q}_4,\mathbf{k}-\mathbf{q}_2)
  }\hskip .2em
  \overline{
    e^{i\mathbf{q}_3\cdot\mathbf{R}_n}\left[z\mathcal{I}-\mathcal{H}\right]_{nl}^{-1} 
    e^{-i\mathbf{q}_4\cdot\mathbf{R}_l}}
  \nonumber \\ &=&
  -\frac{N}{V^2} \sum_{\mathbf{q}_3} \left(f(\mathbf{q}_1-\mathbf{q}_3) - f(q_3) \right)
  \left(f(\mathbf{q}_2-\mathbf{q}_3) - f(q_3) \right)
  \left[\delta_{\mathbf{q}_1,\mathbf{q}_2} +
    \delta_{\mathbf{q}_1-\mathbf{q}_3-\mathbf{k}+\mathbf{q}_2,\mathbf{0}}\right]G(\mathbf{q}_3;z) 
  \nonumber \\ &=&
  -\frac{N}{V^2} \sum_{\mathbf{q}_3} \left(f(\mathbf{q}_1-\mathbf{q}_3) - f(q_3) \right)^2G(\mathbf{q}_3;z)
  \delta_{\mathbf{q}_1,\mathbf{q}_2}
  \nonumber \\ &&
  -\frac{N}{V^2} \left(f(\mathbf{k}-\mathbf{q}_2) - f(\mathbf{q}_1+\mathbf{q}_2-\mathbf{k}) \right)
  \left(f(\mathbf{k}-\mathbf{q}_1) - f(\mathbf{q}_1+\mathbf{q}_2-\mathbf{k}) \right)
  G(\mathbf{q}_1+\mathbf{q}_2-\mathbf{k};z).
\end{eqnarray}
\end{widetext}
In the fourth line of Eq. \eqref{seirr5} we used the following notation
\begin{eqnarray}\label{n2}
  && n_{2}(\mathbf{q}_2-\mathbf{q}_4,\mathbf{k}-\mathbf{q}_2) =
  \sum_{n} 
  \left[e^{-i(\mathbf{q}_2-\mathbf{q}_4)\cdot\mathbf{R}_{n}} -
    \overline{e^{-i(\mathbf{q}_2-\mathbf{q}_4)\cdot\mathbf{R}_{n}}}\right]
  \nonumber \\ && \times
  \sum_{k\neq n} \left[e^{-i\left(\mathbf{k}-\mathbf{q}_2\right)\cdot\mathbf{R}_k} -
    \overline{e^{-i\left(\mathbf{k}-\mathbf{q}_2\right)\cdot\mathbf{R}_k}}\right].
\end{eqnarray}
  
We note that the factorization approximation formulated in Eq. \eqref{seirr5} is essentially the same as the final
factorization approximation of Vogel \textit{et al.} \cite{Vogel2024}. The resulting self-consistent theory,
consisting of Eqs. \eqref{res12}, \eqref{seirr2}, \eqref{renver9}, \eqref{seirr5}, bears strong
resemblance to Leutheusser's theory of random Lorentz gas \cite{Leutheusser}. Notably, it includes
the so-called off-diagonal contributions to self-energy $\Sigma$. 

\section{Discussion: dispersion relation \& Rayleigh scaling}\label{discussion}

Factorization approximation \eqref{seirr5} leaves us with a closed system of equations,
Eqs. \eqref{res12}, \eqref{seirr2}, \eqref{renver9}, \eqref{seirr5}. A self-consistent solution
of these equations will likely be somewhat demanding, due to associated wave-vector integrations.
But, following Vogel and Fuchs \cite{VogelPRL2023} and Vogel \textit{et al.} \cite{Vogel2024}, some qualitative 
results can be obtained without the full numerical solution.

For example, one can derive the dispersion relation, \textit{i.e.} an integral equation for for the
wavevector-dependent speed of sound squared. To this end one notices that the renormalized
wavevector dependent speed of sound $v(k)$ can be expressed in terms of  resolvent $G(\mathbf{k};z)$,
Eq. \eqref{resdef}, 
calculated at $z=0$ \cite{VogelPRL2023,Vogel2024},
\begin{eqnarray}\label{speedren}
  \left(v(k) k \right)^{-2} = - G(\mathbf{k};0).
\end{eqnarray}
Furthermore, at $z=0$ the self-consistent system of equations,
Eqs. \eqref{res12}, \eqref{seirr2}, \eqref{renver9}, \eqref{seirr5}, simplifies considerably.
Specifically, one gets,
\begin{widetext}
\begin{eqnarray}\label{disp1}
  && \left(c(k) \right)^{-2}  + \frac{\rho}{Vk^2 \left(c(k) \right)^4} \sum_{\mathbf{q}_1}
  \left( f(\mathbf{k}-\mathbf{q}_1)-f(q_1) \right)
  \sum_{\mathbf{q}_2} \left[\mathcal{F}_\mathbf{k}(\mathbf{q}_1,\mathbf{q}_2) \right]^{-1}
  \left( f(\mathbf{k}-\mathbf{q}_2) - f(q_2) \right) =
  \left(v(k) \right)^{-2} 
  \\ && \nonumber 
   -  \frac{\rho}{Vk^2 \left(c(k) \right)^4} \sum_{\mathbf{q}_1}
   \left( f(\mathbf{k}-\mathbf{q}_1)-f(q_1) \right)
   \sum_{\mathbf{q}_2} \left[\mathcal{F}_\mathbf{k}(\mathbf{q}_1,\mathbf{q}_2) \right]^{-1}
   \sum_{\mathbf{q}_3} \Sigma^{\mathcal{V}}(\mathbf{q}_2,\mathbf{q}_3,\mathbf{k};0)
  \sum_{\mathbf{q}_4} \left[\mathcal{F}_\mathbf{k}(\mathbf{q}_3,\mathbf{q}_4) \right]^{-1}
  \left( f(\mathbf{k}-\mathbf{q}_4) - f(q_4) \right)
\end{eqnarray}
\begin{eqnarray}\label{disp2}
  \Sigma^{\mathcal{V}}(\mathbf{q}_2,\mathbf{q}_3,\mathbf{k};0) &=& 
  \frac{\rho}{V} \sum_{\mathbf{q}_5} \left(f(\mathbf{q}_2-\mathbf{q}_5) - f(q_5) \right)^2\left(v(q_5) q_5 \right)^{-2}
  \delta_{\mathbf{q}_2,\mathbf{q}_3}
  \\ &+& \nonumber 
  \frac{\rho}{V} \left(f(\mathbf{k}-\mathbf{q}_3) - f(\mathbf{q}_2+\mathbf{q}_3-\mathbf{k}) \right)
  \left(f(\mathbf{k}-\mathbf{q}_2) - f(\mathbf{q}_2+\mathbf{q}_3-\mathbf{k}) \right)
  \left(v(|\mathbf{q}_2+\mathbf{q}_3-\mathbf{k}|) |\mathbf{q}_2+\mathbf{q}_3-\mathbf{k}|\right)^{-2}.
\end{eqnarray}
In Eq. \eqref{disp1} $c(k)$ is the bare (un-renormalized) speed of sound,
$c(k) = \rho^{1/2} \sqrt{f(0)-f(k)}/k$.

Equations (\ref{disp1}-\ref{disp2})  have the form similar to that 
of the corresponding equation derived by Vogel \textit{et al.} (see Eq. (11) in Ref. \cite{Vogel2024}).  
In particular, the second line of Eq. \eqref{disp1} with $\Sigma^\mathcal{V}$ given
by Eq. \eqref{disp2} can be written as a linear operator
$-p^{d-3}\mathcal{C}_{k,p}$ acting on $\left(v^2(p) \right)^{-2}$,
\textit{i.e.} as $-\sum_p p^{d-3}\mathcal{C}_{k,p} \left(v^2(p) \right)^{-2}$, where $\mathcal{C}_{k,p}$ is
given by the following expression
\begin{eqnarray}\label{linop1}
  && \mathcal{C}_{k,p} = \frac{\rho^2}{V^2k^2 \left(c(k) \right)^4} \int d\hat{\mathbf{p}}
   \sum_{\mathbf{q}_1}
  \sum_{\mathbf{q}_2} 
  \sum_{\mathbf{q}_3} 
  \left( f(\mathbf{k}-\mathbf{q}_1)-f(q_1) \right)
  \left\{
  \left[\mathcal{F}_\mathbf{k}(\mathbf{q}_1,\mathbf{q}_2) \right]^{-1}
    \left(f(\mathbf{q}_2-\mathbf{p}) - f(p) \right)^2 
    \left[\mathcal{F}_\mathbf{k}(\mathbf{q}_2,\mathbf{q}_3) \right]^{-1}
    \right.
    \nonumber \\ && \left.
    +  \left[\mathcal{F}_\mathbf{k}(\mathbf{q}_1,\mathbf{q}_2) \right]^{-1}
    \left(f(\mathbf{q}_2-\mathbf{p}) - f(\mathbf{p}) \right)
    \left(f(\mathbf{k}-\mathbf{q}_2) - f(\mathbf{p}) \right)
    \left[\mathcal{F}_\mathbf{k}(\mathbf{k}+\mathbf{p}-\mathbf{q}_2,\mathbf{q}_3) \right]^{-1}
    \right\}
  \left( f(\mathbf{k}-\mathbf{q}_3) - f(q_3) \right)
\end{eqnarray}
\end{widetext}
Linear operator $\mathcal{C}_{k,p}$ is referred to as the stability matrix in Ref. \cite{Vogel2024}. 
The main difference between our expression and that of Vogel \textit{et al.}
originates from the fact that the authors of Ref. \cite{Vogel2024}
used an approximation that made the analog of our kernel
$\mathcal{F}_{\mathbf{k}}(\mathbf{q}_1,\mathbf{q}_2)$ diagonal in $\mathbf{q}_1$, $\mathbf{q}_2$.
However, both equations lead to similar conclusions.
Since in the large density limit the bare speed of sound  
increases as $\rho^{1/2}$ and $ \left[\mathcal{F}_\mathbf{k}(\mathbf{q}_4,\mathbf{q}_1) \right]^{-1}$
decreases as $\rho^{-1}$, 
one can argue that in that limit linear operator $q^{d-3}\mathcal{C}_{k,p}$ in the second line of
Eq. \eqref{disp1} can be neglected. The inverse of speed of sound,  $\left(v(p) \right)^{-2}$,
is then given by the positive definite
quantity at the left-hand-side of Eq. \eqref{disp1}. On physical grounds we expect that the magnitude of 
$\mathcal{C}_{k,p}$ increases with decreasing density. The inverse of speed of sound,  $\left(v(p) \right)^{-2}$,
is then determined by
the inverse of $\mathcal{I}-p^{d-3}\mathcal{C}_{k,p}$ acting on the left-hand-side of Eq. \eqref{disp1}.
For small enough density it may happen that largest eigenvalue of $p^{d-3}\mathcal{C}_{k,p}$ exceeds unity.
At this density the inverse of the speed of sound diverges, \textit{i.e.} the speed of sound vanishes.
This corresponds to an un-jamming transition \cite{Vogel2024}.
For the standard Gaussian form of the spring constant
that we implicitly assume in this paper, on physical grounds we do not expect un-jamming transition.
However, an explicit numerical solution of Eqs. (\ref{disp1}-\ref{disp2}) is needed in order to check whether
this expectation is fulfilled within our approximate self-consistent theory.

Furthermore, a direct evaluation shows that our linear operator vanishes at $k=0=p$, like
the corresponding operator of Vogel \textit{et al.} \cite{Vogel2024}.
This fact was the crucial feature that allowed
Vogel \textit{et al.} to show that their theory reproduces the Rayleigh scattering scaling of
the sound attenuation coefficient on the wavevector.  Thus, the analysis of Vogel \textit{et al.},
presented in Sec. III.B of their paper \cite{Vogel2024}, 
can be repeated for our theory, leading to the Rayleigh scaling and an
expression for the magnitude of the sound attenuation coefficient.

\section{Concluding remarks}\label{final}

Recent results of Baumg\"artel \textit{et al.} \cite{Baumgaertel2024}
suggest that for the ERM model, the contribution of the self-energy to the dispersion relation is rather small,
at least for high densities. This suggests that future work should concentrate on extending
the present theory to more realistic models of amorphous solids. The are two possible problems that may
impede such an extension. First, one would need to assume a more realistic distribution of particle
configurations, which incorporates correlations between positions of the particles that are present in
actual amorphous solids. Such correlations would make projection operators more complicated. In addition, the
presence of higher-order correlations would likely force one to approximate them through products of
pair correlations. Second, it is not clear whether it is necessary/important to take into account the
non-equilibrium character of the probability distribution of particle configurations and if yes, how to
deal with the absence of many equilibrium relations that are usually used in projection operator
manipulations. We hope to address both problems in the near future. 

\section*{Acknowledgments}

I thank Florian Vogel and Matthias Fuchs for inspiring discussions
on approximate theories of sound propagation in amorphous materials. 
I also thank them and Elijah Flenner for comments on the manuscript.
This work was started in 2023, when I was visiting \'{E}cole Normale Sup\'{e}rieure de Paris.
I thank my colleagues there for their hospitality.
The visit to ENS was partially supported by the Simons Foundation Grant 454955 (to Francesco Zamponi).
I gratefully acknowledge the support of NSF Grant No.~CHE 2154241. 

\begin{widetext}
  \appendix
  
  \section{Projection operator analysis of the resolvent}\label{appA}
  
  We start by inserting identity written as $\mathcal{P}_j+\mathcal{Q}_j$ between $\mathcal{H}_{ij}$ and
  $\left[z\mathcal{I}-\mathcal{H}\right]^{-1}_{jk}$ at the right-hand-side of Eq. \eqref{res1} of the main text
  and we get
  \begin{eqnarray}\label{res2}
    zG(\mathbf{k};z) - 1 =
    \overline{ \sum_{j} n_1(-\mathbf{k}) \mathcal{H}_{1j} n_j(\mathbf{k}) } 
    G(\mathbf{k};z)
    + \overline{\sum_{j,k} n_1(-\mathbf{k}) \mathcal{H}_{1j}\mathcal{Q}_j
      \left[z\mathcal{I}-\mathcal{H}\right]_{jk}^{-1} n_k(\mathbf{k})}.
  \end{eqnarray}
  
  The first term at the right-hand-side of Eq. \eqref{res2} is a product of
  a matrix element of the Hessian and the resolvent. The matrix element of the Hessian
  plays the role of the frequency matrix for the resolvent. It reads
  \begin{eqnarray}\label{PHP}
    && \overline{ \sum_{j} n_1(-\mathbf{k}) \mathcal{H}_{1j} n_j(\mathbf{k}) }
    = \rho \left(f(0)-f(k)\right),
  \end{eqnarray}
  where $\rho=N/V$ is the number density.
  
  To proceed, it is convenient to introduce an auxiliary function $g_i(\mathbf{k};z)$,
  \begin{eqnarray}\label{res3}
    g_i(\mathbf{k};z) = 
    \mathcal{Q}_i \sum_k 
    \left[z\mathcal{I}-\mathcal{H}\right]_{ik}^{-1} n_j(\mathbf{k}).
  \end{eqnarray}
  
  We substitute Eq. \eqref{PHP} into Eq. \eqref{res2} and then, following Ref. \cite{GSPRE2023},
  we rewrite the resulting expression as follows,
  \begin{eqnarray}\label{res4}
    G(\mathbf{k};z) = \left[\rho \left(f(0)-f(k)\right) \right]^{-1} \left[zG(\mathbf{k};z) - 1\right]
    - \left[\rho \left(f(0)-f(k)\right) \right]^{-1}
    \overline{ \sum_{j} n_1(-\mathbf{k})  \mathcal{H}_{1j}\mathcal{Q}_j g_j(\mathbf{k};z)}.
  \end{eqnarray}
  
  Then, we consider $zg_i$,
  \begin{eqnarray}\label{res5}
    z g_i(\mathbf{k};z)  &=&  z\mathcal{Q}_i \sum_k 
    \left[z\mathcal{I}-\mathcal{H}\right]_{ik}^{-1}n_k(\mathbf{k})
    \sum_{j,k} \mathcal{Q}_i \mathcal{H}_{ij}
    \left(\mathcal{P}_j+\mathcal{Q}_j\right)
    \left[z\mathcal{I}-\mathcal{H}\right]_{jk}^{-1}n_k(\mathbf{k})
    \\ \nonumber &=& 
    \sum_{j} \mathcal{Q}_i \mathcal{H}_{ij} n_j(\mathbf{k})
    G(\mathbf{k};z)
    +  \sum_{j} \mathcal{Q}_i \mathcal{H}_{ij}
    \mathcal{Q}_j g_j(\mathbf{k};z) 
  \end{eqnarray}
  
  Next, we substitute the right-hand-side of Eq. \eqref{res4} for the resolvent in Eq. \eqref{res5} and we get
  \begin{eqnarray}\label{res6}
    z g_i(\mathbf{k};z)
    &=& \sum_{j} \mathcal{Q}_i \mathcal{H}_{ij} n_j(\mathbf{k})
    \left[\rho \left(f(0)-f(k)\right) \right]^{-1} \left[zG(\mathbf{k};z) - 1\right]
    \nonumber \\ &&
    - \sum_{j} \mathcal{Q}_i \mathcal{H}_{ij} n_j(\mathbf{k}) \left[\rho \left(f(0)-f(k)\right) \right]^{-1}
    \mathcal{P} \sum_m n_j(-\mathbf{k})\mathcal{H}_{jm}\mathcal{Q}_m g_m(\mathbf{k};z)
    +  \sum_{j} \mathcal{Q}_i \mathcal{H}_{ij}
    \mathcal{Q}_j g_j(\mathbf{k};z),
  \end{eqnarray}
  where we used $\mathcal{P}$ defined in Eq. \eqref{Pdef}.

  Finally, we solve Eq. \eqref{res6} for $g_i(\mathbf{k};z)$, we substitute the result into \eqref{res2} and we get,
  formula \eqref{res8} of the main text.
  
\section{Second projection operator}\label{appB}

The second projection operator projects on the product of the single site density of site
$j$ and the fluctuation of the collective density of all other sites,
\begin{eqnarray}\label{APj2def}
  \mathcal{P}_{j2}^{\mathbf{q}_1} f &=& \sum_{\mathbf{q}_2}
  n_{j2}(\mathbf{q}_1,\mathbf{k}-\mathbf{q}_1)
  \overline{n_{j2}(-\mathbf{q}_1,-\mathbf{k}+\mathbf{q}_1)
    n_{j2}(\mathbf{q}_2,\mathbf{k}-\mathbf{q}_2)}^{-1}
  \overline{ n_{j2}(-\mathbf{q}_2,-\mathbf{k}+\mathbf{q}_2) f}.
\end{eqnarray}

The absence of interparticle correlations allows us to explicitly evaluate the normalization factor
in definition \eqref{APj2def}, which is given by the following three-particle average,
\begin{eqnarray}\label{Pj2norm1}
  && \overline{n_{j2}(-\mathbf{q}_1,-\mathbf{k}+\mathbf{q}_1)
    n_{j2}(\mathbf{q}_2,\mathbf{k}-\mathbf{q}_2)}
  \nonumber \\ &=&
  \overline{e^{i\mathbf{q}_1\cdot\mathbf{R}_j} \left[\sum_{l\neq j} e^{i\left(\mathbf{k}-\mathbf{q}_1\right)\cdot\mathbf{R}_l} -
      \overline{\sum_{l\neq j} e^{i\left(\mathbf{k}-\mathbf{q}_1\right)\cdot\mathbf{R}_l}}\right]
    e^{-i\mathbf{q}_2\cdot\mathbf{R}_j} \left[\sum_{m\neq j} e^{-i\left(\mathbf{k}-\mathbf{q}_2\right)\cdot\mathbf{R}_m} -
      \overline{\sum_{m\neq j} e^{-i\left(\mathbf{k}-\mathbf{q}_2\right)\cdot\mathbf{R}_m}}\right]}
  \\ \nonumber &=&
  \delta_{\mathbf{q}_1,\mathbf{q}_2} \overline{\sum_{l\neq j} \left[e^{i\left(\mathbf{k}-\mathbf{q}_1\right)\cdot\mathbf{R}_l} -
      \overline{e^{i\left(\mathbf{k}-\mathbf{q}_1\right)\cdot\mathbf{R}_l}}\right]
    \left[e^{-i\left(\mathbf{k}-\mathbf{q}_1\right)\cdot\mathbf{R}_l} -
      \overline{e^{-i\left(\mathbf{k}-\mathbf{q}_1\right)\cdot\mathbf{R}_l}}\right]} =
  (N-1) \delta_{\mathbf{q}_1,\mathbf{q}_2} \left(1-\delta_{\mathbf{q}_1,\mathbf{k}}\right) \approx
  N\delta_{\mathbf{q}_1,\mathbf{q}_2}.
\end{eqnarray}

Equation \eqref{Pj2norm1} allows us to re-write definition \eqref{APj2def} of the second projection operator
as follows,
\begin{eqnarray}\label{APj2def2}
  \mathcal{P}_{j2}^{\mathbf{q}_1} f
  =
  \frac{1}{N}n_{j2}(\mathbf{q}_1,\mathbf{k}-\mathbf{q}_1)
  \overline{ n_{j2}(-\mathbf{q}_1,-\mathbf{k}+\mathbf{q}_1) f}.
  \nonumber \\
\end{eqnarray}

\section{Auxiliary quantity $\mathcal{V}_0(\mathbf{q}_1,\mathbf{k}) =
  \lim_{z\to\infty} z\mathcal{V}(\mathbf{q}_1,\mathbf{k};z)$}\label{appC}

Auxiliary quantity $\mathcal{V}_0(\mathbf{q}_1,\mathbf{k}) =
\lim_{z\to\infty} z\mathcal{V}(\mathbf{q}_1,\mathbf{k};z)$ allows us to write Eq. \eqref{renver3} in
a relatively compact way. This quantity can be evaluated as follows,
\begin{eqnarray}\label{Aver01a}
  \mathcal{V}_0(\mathbf{q}_1,\mathbf{k}) &=&
  - \lim_{z\to\infty} 
  \sum_{l} \sum_{\mathbf{q}_2}
  \frac{1}{V}  z \overline{n_{j2}(-\mathbf{q}_1,\mathbf{q}_1-\mathbf{k})
    \left[z\mathcal{I} - \mathcal{H}^\Sigma \right]^{-1}_{jl}
    n_{l2}(\mathbf{q}_2,\mathbf{k}-\mathbf{q}_2)}
  \left( f(\mathbf{k}-\mathbf{q}_2) - f(q_2) \right)
  \nonumber \\ &=& 
  -  \sum_{l} \sum_{\mathbf{q}_2}
  \frac{1}{V}  \overline{n_{j2}(-\mathbf{q}_1,\mathbf{q}_1-\mathbf{k})\delta_{jl}
    n_{l2}(\mathbf{q}_2,\mathbf{k}-\mathbf{q}_2)}
  \left( f(\mathbf{k}-\mathbf{q}_2) - f(q_2) \right).  
\end{eqnarray}
Using Eq. \eqref{Pj2norm1} we re-write the last expression as follows,
\begin{eqnarray}\label{Aver01b}
  && - \sum_{\mathbf{q}_2}
  \frac{1}{V}  \overline{n_{j2}(-\mathbf{q}_1,\mathbf{q}_1-\mathbf{k})
    n_{j2}(\mathbf{q}_2,\mathbf{k}-\mathbf{q}_2)}
  \left( f(\mathbf{k}-\mathbf{q}_2) - f(q_2) \right) =
  - \sum_{\mathbf{q}_2}
  \frac{1}{V}  N\delta_{\mathbf{q}_1,\mathbf{q}_2}  \left( f(\mathbf{k}-\mathbf{q}_2) - f(q_2) \right)
  \nonumber \\ &=&
  -\rho \left( f(\mathbf{k}-\mathbf{q}_1) - f(q_1) \right).
\end{eqnarray}

\section{Projection operator analysis of the vertex function}\label{appD}

We start by inserting identity written as
$\sum_{\mathbf{q}_3}\mathcal{P}_{n2}^{\mathbf{q}_3}+\mathcal{Q}_{n2}$
between $\mathcal{H}^\Sigma_{jn}$ and $\left[z\mathcal{I} - \mathcal{H}^\Sigma \right]^{-1}_{nl}$ 
at the right-hand-side of Eq. \eqref{renver3} of the main text 
and we get (note the analogy with Eq. \eqref{res2}),
\begin{eqnarray}\label{renver4}
  && z\mathcal{V}(\mathbf{q}_1,\mathbf{k};z) - \mathcal{V}_0(\mathbf{q}_1,\mathbf{k}) =
   \frac{1}{N} \sum_{\mathbf{q}_3}
   \overline{\sum_{n} n_{j2}(-\mathbf{q}_1,-\mathbf{k}+\mathbf{q}_1)
     \mathcal{H}^\Sigma_{jn} 
     n_{n2}(\mathbf{q}_3,\mathbf{k}-\mathbf{q}_3)}
   \mathcal{V}(\mathbf{q}_3,\mathbf{k};z) 
  \nonumber \\ &&
  - \frac{1}{V} 
  \sum_{\mathbf{q}_2}
  \overline{\sum_{n,l} n_{j2}(-\mathbf{q}_1,-\mathbf{k}+\mathbf{q}_1)
    \mathcal{H}^\Sigma_{jn}
  \mathcal{Q}_{n2}
  \left[z\mathcal{I} - \mathcal{H}^\Sigma \right]^{-1}_{nl} 
  n_{l2}(\mathbf{q}_2,\mathbf{k}-\mathbf{q}_2)}
  \left( f(\mathbf{k}-\mathbf{q}_2) - f(-\mathbf{q}_2) \right)
\end{eqnarray}

The first term at the right-hand-side of Eq. \eqref{renver4} is a convolution of the
vertex function with a matrix element of the projected Hessian. The latter quantity
is essentially the frequency matrix for the vertex function,
\begin{eqnarray}\label{2freqmatdef}
  \mathcal{F}_{\mathbf{k}}(\mathbf{q}_1,\mathbf{q}_2)
  &=&  \frac{1}{N} 
  \overline{\sum_{n} n_{j2}(-\mathbf{q}_1,-\mathbf{k}+\mathbf{q}_1)
    \mathcal{H}^\Sigma_{jn} n_{n2}(\mathbf{q}_2,\mathbf{k}-\mathbf{q}_2)} =
  \frac{1}{N} 
  \overline{\sum_{n} n_{j2}(-\mathbf{q}_1,-\mathbf{k}+\mathbf{q}_1)
    \mathcal{H}_{jn} n_{n2}(\mathbf{q}_2,\mathbf{k}-\mathbf{q}_2)}
  \nonumber \\ &&
  -  \frac{1}{N}
  \sum_{k,n} \overline{n_{j2}(-\mathbf{q}_1,-\mathbf{k}+\mathbf{q}_1) \mathcal{H}_{jk} n_k(\mathbf{k})}
  \left[\rho \left(f(0)-f(k)\right) \right]^{-1}
  \overline{n_k(-\mathbf{k})\mathcal{H}_{kn} n_{n2}(\mathbf{q}_2,\mathbf{k}-\mathbf{q}_2)}.
\end{eqnarray}

To evaluate the frequency matrix $\mathcal{F}_{\mathbf{k}}(\mathbf{q}_1,\mathbf{q}_2)$ we start with analyzing
$\sum_n \mathcal{H}_{jn} n_{n2}(\mathbf{q}_2,\mathbf{k}-\mathbf{q}_2)$. A series of somewhat tedious
but straightforward steps allows us to re-write this quantity in terms of a three-particle
density and densities $n_j$ and $n_{j2}$,
\begin{eqnarray}\label{2freqmat1}
  && \sum_n \mathcal{H}_{jn}
  n_{n2}(\mathbf{q}_2,\mathbf{k}-\mathbf{q}_2) =
  \frac{1}{V}\sum_{\mathbf{q}_3}\left[ f(\mathbf{q}_2-\mathbf{q}_3) - f(-\mathbf{q}_3)  \right]
    e^{-i\mathbf{q}_3\cdot\mathbf{R}_j}
  \sum_{n\neq j} \sum_{k\neq j,k\neq n} e^{-i(\mathbf{q}_2-\mathbf{q}_3)\cdot\mathbf{R}_{n}}
  \left[e^{-i\left(\mathbf{k}-\mathbf{q}_2\right)\cdot\mathbf{R}_k} -
    \overline{e^{-i\left(\mathbf{k}-\mathbf{q}_2\right)\cdot\mathbf{R}_k}}\right]
  \nonumber \\ &&
  +  \frac{1}{V}\sum_{\mathbf{q}_3}f(\mathbf{q}_2-\mathbf{q}_3)
  \left[n_{j2}(\mathbf{q}_3,\mathbf{k}-\mathbf{q}_3)
    + n_j(\mathbf{q}_3) (N-1) \delta_{\mathbf{k},\mathbf{q}_3}
    - n_{j2}(\mathbf{q}_3,\mathbf{q}_2-\mathbf{q}_3)\delta_{\mathbf{k},\mathbf{q}_2}
    - n_j(\mathbf{q}_3)(N-1)\delta_{\mathbf{q}_2,\mathbf{q}_3}\delta_{\mathbf{k},\mathbf{q}_2}
    \right]
  \nonumber \\ &&
  - \frac{1}{V}\sum_{\mathbf{q}_3}  f(-\mathbf{q}_3)
  \left[ n_{j2}(\mathbf{k}-\mathbf{q}_2+\mathbf{q}_3,\mathbf{q}_2-\mathbf{q}_3) +
    n_j(\mathbf{k}-\mathbf{q}_2+\mathbf{q}_3)(N-1)\delta_{\mathbf{q}_2,\mathbf{q}_3}
    \right. \nonumber \\ && \left.
    - n_{j2}(\mathbf{q}_3,\mathbf{q}_2-\mathbf{q}_3)\delta_{\mathbf{k},\mathbf{q}_2}
    - n_j(\mathbf{q}_3)(N-1)\delta_{\mathbf{q}_2,\mathbf{q}_3}\delta_{\mathbf{k},\mathbf{q}_2}
    \right].
\end{eqnarray}

Frequency matrix \eqref{2freqmatdef} consists of two terms, a matrix element of the Hessian,
$N^{-1} \overline{\sum_n n_{2j}\mathcal{H}_{jn} n_{2n}}$, and
a subtraction term. Using Eq. \eqref{2freqmat1} and the fact that quantities $n_j$ and $n_{j2}$ are orthogonal,
$\overline{n_j n_{j2}} = 0$, we can conclude that the dominant terms contributing to the
matrix element of the Hessian are
\begin{eqnarray}\label{2freqmat2}
  && \frac{1}{NV}\sum_{\mathbf{q}_3}\left[ f(\mathbf{q}_2-\mathbf{q}_3) - f(-\mathbf{q}_3)  \right]
  \overline{ n_{j2}(-\mathbf{q}_1,-\mathbf{k}+\mathbf{q}_1)
    e^{-i\mathbf{q}_3\cdot\mathbf{R}_j}
    \sum_{n\neq j} \sum_{k\neq j,k\neq n} e^{-i(\mathbf{q}_2-\mathbf{q}_3)\cdot\mathbf{R}_{n}}
    \left[e^{-i\left(\mathbf{k}-\mathbf{q}_2\right)\cdot\mathbf{R}_k} -
      \overline{e^{-i\left(\mathbf{k}-\mathbf{q}_2\right)\cdot\mathbf{R}_k}}\right]}
  \nonumber \\ &=&
  n \left[f(0) - f(q_2)\right]\delta_{\mathbf{q}_1,\mathbf{q}_2},
\end{eqnarray}
\begin{eqnarray}\label{2freqmat3}
  \frac{1}{NV}\sum_{\mathbf{q}_3}f(\mathbf{q}_2-\mathbf{q}_3)
  \overline{ n_{j2}(-\mathbf{q}_1,-\mathbf{k}+\mathbf{q}_1) n_{j2}(\mathbf{q}_3,\mathbf{k}-\mathbf{q}_3)}
  = \frac{1}{V}f(\mathbf{q}_2-\mathbf{q}_1),
\end{eqnarray}
\begin{eqnarray}\label{2freqmat4}
  -\frac{1}{NV}\sum_{\mathbf{q}_3}  f(-\mathbf{q}_3)
  \overline{ n_{j2}(-\mathbf{q}_1,-\mathbf{k}+\mathbf{q}_1)
    n_{j2}(\mathbf{k}-\mathbf{q}_2+\mathbf{q}_3,\mathbf{q}_2-\mathbf{q}_3) }
  =  - \frac{1}{V}f(\mathbf{k}-\mathbf{q}_2-\mathbf{q}_2).
\end{eqnarray}

To calculate the subtraction term we need off-diagonal matrix element of the Hessian,
$\overline{\sum_n n_{k}\mathcal{H}_{kn} n_{2n}}$. Using Eq. \eqref{2freqmat1} and the fact that
quantities $n_j$ and $n_{j2}$ are orthogonal we can conclude that the dominant terms contributing to this
matrix element are
\begin{eqnarray}\label{2freqmat5}
  \frac{1}{V} \sum_{\mathbf{q}_3}f(\mathbf{q}_2-\mathbf{q}_3)
  \overline{n_k(-\mathbf{k}) n_k(\mathbf{q}_3)}(N-1)\delta_{\mathbf{k},\mathbf{q}_3} =
  \frac{N}{V} f(\mathbf{q}_2-\mathbf{k}),
\end{eqnarray}
\begin{eqnarray}\label{2freqmat6}
  -\frac{1}{V} \sum_{\mathbf{q}_3}  f(-\mathbf{q}_3)
  \overline{n_k(-\mathbf{k}) n_j(\mathbf{k}-\mathbf{q}_2+\mathbf{q}_3)}(N-1)\delta_{\mathbf{q}_2,\mathbf{q}_3}
  =-\frac{N}{V}f(q_2).
\end{eqnarray}
  
Combining Eqs. (\ref{2freqmat2}-\ref{2freqmat6}) we obtain
\begin{eqnarray}\label{Hirr6}
  \mathcal{F}_{\mathbf{k}}(\mathbf{q}_1,\mathbf{q}_3)
    &=& \rho \left(f(0)-f(q_3)\right) \delta_{\mathbf{q}_1,\mathbf{q}_3} 
    \nonumber \\ &&
    + \frac{1}{V}\left(f(\mathbf{q}_3-\mathbf{q}_1)-f(\mathbf{k}-\mathbf{q}_1-\mathbf{q}_3)\right)
    - \frac{1}{V} \frac{ \left(f(-\mathbf{q}_1+\mathbf{k})- f(q_1) \right)
      \left(f(\mathbf{q}_3-\mathbf{k})- f(q_3) \right)}{\left(f(0)-f(k)\right) }
  \nonumber \\
\end{eqnarray}

To proceed with the analysis of Eq. \eqref{renver4}, it is convenient to introduce an auxiliary function
$v_i(\mathbf{k};z)$ (note the analogy with Eq. \eqref{res3}),
\begin{eqnarray}\label{vi}
  v_i(\mathbf{k};z) = -\frac{1}{V}\sum_{\mathbf{q}_2} \sum_{l} \mathcal{Q}_{i2} 
  \left[z\mathcal{I}-\mathcal{H}^\Sigma\right]_{il}^{-1} 
  n_{l2}(\mathbf{q}_2,\mathbf{k}-\mathbf{q}_2)
      \left( f(\mathbf{k}-\mathbf{q}_2) - f(-\mathbf{q}_2) \right).
\end{eqnarray}

We substitute $\mathcal{F}$ defined in Eq. \eqref{Hirr6} into Eq. \eqref{renver4} and then,
following the derivation of the self-energy in Sec. \ref{selfenergy}, we rewrite the resulting expression
to obtain (note the analogy with Eq. \eqref{res4}),
\begin{eqnarray}\label{renver5}
  && \mathcal{V}(\mathbf{q}_1,\mathbf{k};z)
  = \sum_{\mathbf{q}_3} \left[\mathcal{F}_\mathbf{k}(\mathbf{q}_1,\mathbf{q}_3) \right]^{-1}
  \left[z\mathcal{V}(\mathbf{q}_3,\mathbf{k};z) - \mathcal{V}_0(\mathbf{q}_3,\mathbf{k})\right]
  \nonumber \\ &&
  - \sum_{\mathbf{q}_3} \left[\mathcal{F}_\mathbf{k}(\mathbf{q}_1,\mathbf{q}_3) \right]^{-1}
  \overline{\sum_{n} n_{j2}(-\mathbf{q}_3,-\mathbf{k}+\mathbf{q}_3)
    \mathcal{H}^\Sigma_{jn}
    \mathcal{Q}_{n2} v_n(\mathbf{k};z)}.
\end{eqnarray}

Then we consider $zv_i$ (note the analogy with Eq. \eqref{res5}),
\begin{eqnarray}\label{renver6}
  && z v_i(\mathbf{k};z) =
  -\frac{1}{V}\sum_{\mathbf{q}_2} \sum_{j,l} \mathcal{Q}_{i2} \mathcal{H}^\Sigma_{ij}
  \left[z\mathcal{I}-\mathcal{H}^\Sigma\right]_{jl}^{-1} 
  n_{l2}(\mathbf{q}_2,\mathbf{k}-\mathbf{q}_2)\left( f(\mathbf{k}-\mathbf{q}_2) - f(-\mathbf{q}_2) \right)
  \\ \nonumber &=& 
  -\frac{1}{V} \sum_{\mathbf{q}_2} \sum_{j,l} \mathcal{Q}_{i2} \mathcal{H}^\Sigma_{ij}
  \left(\sum_{\mathbf{q}_1}\mathcal{P}^{\mathbf{q}_1}_{j2}+\mathcal{Q}_{j2}\right)
  \left[z\mathcal{I}-\mathcal{H}^\Sigma\right]_{jl}^{-1} 
  n_{l2}(\mathbf{q}_2,\mathbf{k}-\mathbf{q}_2)\left( f(\mathbf{k}-\mathbf{q}_2) - f(-\mathbf{q}_2) \right)
  \\ \nonumber &=& 
  \frac{1}{N} \sum_{\mathbf{q}_1} \sum_{j} \mathcal{Q}_{i2} \mathcal{H}^\Sigma_{ij}  
  n_{j2}(\mathbf{q}_1,\mathbf{k}-\mathbf{q}_1)
  \mathcal{V}(\mathbf{q}_1.\mathbf{k};z)
  +  \sum_{j} \mathcal{Q}_{i2} \mathcal{H}^\Sigma_{ij}
  \mathcal{Q}_{j2} v_j(\mathbf{k};z).
\end{eqnarray}

Next, we substitute the right-hand-side of Eq. \eqref{renver5}
for the vertex function in Eq. \eqref{renver6} and we get
(note the analogy with Eq. \eqref{res6})
\begin{eqnarray}\label{renver7}
  && z v_i(\mathbf{k};z) =  \frac{1}{N} \sum_{\mathbf{q}_1} \sum_{j} \mathcal{Q}_{i2} \mathcal{H}^\Sigma_{ij}  
  n_{j2}(\mathbf{q}_1,\mathbf{k}-\mathbf{q}_1)
  \sum_{\mathbf{q}_3} \left[\mathcal{F}_\mathbf{k}(\mathbf{q}_1,\mathbf{q}_3) \right]^{-1}
  \left[z\mathcal{V}(\mathbf{q}_3,\mathbf{k};z) - \mathcal{V}_0(\mathbf{q}_3,\mathbf{k})\right]
  \nonumber \\ &&
  - \frac{1}{N} \sum_{\mathbf{q}_1}  \sum_{j} \mathcal{Q}_{i2} \mathcal{H}^\Sigma_{ij}  
  n_{j2}(\mathbf{q}_1,\mathbf{k}-\mathbf{q}_1)
  \sum_{\mathbf{q}_3} \left[\mathcal{F}_\mathbf{k}(\mathbf{q}_1,\mathbf{q}_3) \right]^{-1}
  \overline{\sum_{n} n_{j2}(-\mathbf{q}_3,-\mathbf{k}+\mathbf{q}_3)
    \mathcal{H}^\Sigma_{jn}
    \mathcal{Q}_{n2} v_n(\mathbf{k};z)}
  \nonumber \\ &&
  +  \sum_{j} \mathcal{Q}_{i2} \mathcal{H}^\Sigma_{ij}
  \mathcal{Q}_{j2} v_j(\mathbf{k};z).
\end{eqnarray}
To write down the solution of the above equation for $v_i$ it is convenient to define the second
projected Hessian (note the analogy with Eq. \eqref{Hirr1}),
\begin{eqnarray}\label{Hirr2}
  \mathcal{H}^{\mathcal{V}}_{ij} =
  \mathcal{Q}_{i2} \mathcal{H}^\Sigma_{ij} \mathcal{Q}_{j2} -
  \frac{1}{N} \sum_{\mathbf{q}_1,\mathbf{q}_3} \sum_{k} \mathcal{Q}_{i2} \mathcal{H}^\Sigma_{ik}  
  n_{k2}(\mathbf{q}_1,\mathbf{k}-\mathbf{q}_1)
  \left[\mathcal{F}_\mathbf{k}(\mathbf{q}_1,\mathbf{q}_3) \right]^{-1}
  \mathcal{P}
  n_{k2}(-\mathbf{q}_3,-\mathbf{k}+\mathbf{q}_3)
  \mathcal{H}^\Sigma_{kj}
  \mathcal{Q}_{j2}.
\end{eqnarray}

Finally, with the help of Eq. \eqref{Hirr2} we write down the solution of  Eq. \eqref{renver7} for $v_i(\mathbf{k};z)$,
we substitute it into Eq. \eqref{renver4} and we get Eq. \eqref{renver8} of the main text.

\section{Three-particle density that determines $\Sigma^\mathcal{V}$}\label{appE}

Using Eq. \eqref{2freqmat1} in Eq. \eqref{vertex2} of the main text we see that the only non-vanishing
contribution to $\sum_m \mathcal{Q}_{l2}\mathcal{Q}_l \mathcal{H}_{lm}
  n_{m2}(\mathbf{q}_2,\mathbf{k}-\mathbf{q}_2)$ originates from 
\begin{eqnarray}\label{vertex5}
  \frac{1}{V}\sum_{\mathbf{q}_3}\left[ f(\mathbf{q}_2-\mathbf{q}_3) - f(-\mathbf{q}_3)  \right]
     \mathcal{Q}_{l2}\mathcal{Q}_l  e^{-i\mathbf{q}_3\cdot\mathbf{R}_l}
  \sum_{n\neq l} \sum_{k\neq l,k\neq n} e^{-i(\mathbf{q}_2-\mathbf{q}_3)\cdot\mathbf{R}_{n}}
  \left[e^{-i\left(\mathbf{k}-\mathbf{q}_2\right)\cdot\mathbf{R}_k} -
    \overline{e^{-i\left(\mathbf{k}-\mathbf{q}_2\right)\cdot\mathbf{R}_k}}\right].
\end{eqnarray}

Due to the subtraction term in the square bracket at the right-hand-side of expression \eqref{vertex5}
orthogonal projection $\mathcal{Q}_l$ does not contribute, and thus we are left with
\begin{eqnarray}\label{vertex6}
  && \frac{1}{V}\sum_{\mathbf{q}_3}\left[ f(\mathbf{q}_2-\mathbf{q}_3) - f(-\mathbf{q}_3)  \right]
  \mathcal{Q}_{l2} e^{-i\mathbf{q}_3\cdot\mathbf{R}_l}
  \sum_{n\neq l} \sum_{k\neq l,k\neq n} e^{-i(\mathbf{q}_2-\mathbf{q}_3)\cdot\mathbf{R}_{n}}
  \left[e^{-i\left(\mathbf{k}-\mathbf{q}_2\right)\cdot\mathbf{R}_k} -
    \overline{e^{-i\left(\mathbf{k}-\mathbf{q}_2\right)\cdot\mathbf{R}_k}}\right]
  \nonumber \\ &=&
  \frac{1}{V}\sum_{\mathbf{q}_3}\left[ f(\mathbf{q}_2-\mathbf{q}_3) - f(-\mathbf{q}_3)  \right]
     e^{-i\mathbf{q}_3\cdot\mathbf{R}_l}
  \sum_{n\neq l} \sum_{k\neq l,k\neq n} e^{-i(\mathbf{q}_2-\mathbf{q}_3)\cdot\mathbf{R}_{n}}
  \left[e^{-i\left(\mathbf{k}-\mathbf{q}_2\right)\cdot\mathbf{R}_k} -
    \overline{e^{-i\left(\mathbf{k}-\mathbf{q}_2\right)\cdot\mathbf{R}_k}}\right] 
  \nonumber \\ &&
  - \frac{1}{V}\sum_{\mathbf{q}_3}\left[ f(\mathbf{q}_2-\mathbf{q}_3) - f(-\mathbf{q}_3)  \right]
     \sum_{\mathbf{q}_4} \mathcal{P}_{l2}^{\mathbf{q}_4} e^{-i\mathbf{q}_3\cdot\mathbf{R}_l}
  \sum_{n\neq l} \sum_{k\neq l,k\neq n} e^{-i(\mathbf{q}_2-\mathbf{q}_3)\cdot\mathbf{R}_{n}}
  \left[e^{-i\left(\mathbf{k}-\mathbf{q}_2\right)\cdot\mathbf{R}_k} -
    \overline{e^{-i\left(\mathbf{k}-\mathbf{q}_2\right)\cdot\mathbf{R}_k}}\right]. 
\end{eqnarray}

Using definition \eqref{Pj2def2} of projection operator $\mathcal{P}_{l2}^{\mathbf{q}_4}$ we get
the following result for the second term at the right-hand-side of Eq. \eqref{vertex6}
\begin{eqnarray}\label{vertex7}
  - \frac{1}{V}\sum_{\mathbf{q}_3}\left[ f(\mathbf{q}_2-\mathbf{q}_3) - f(-\mathbf{q}_3)  \right]
  n_{l2}(\mathbf{q}_3,\mathbf{k}-\mathbf{q}_3)
     (N-1) \delta_{\mathbf{q}_2,\mathbf{q}_3}.
\end{eqnarray}

Combining expression \eqref{vertex6} with the first term at the right-hand-side of Eq. \eqref{vertex6}
we get the following formula for $\sum_m \mathcal{Q}_{l2}\mathcal{Q}_l \mathcal{H}_{lm}
n_{m2}(\mathbf{q}_2,\mathbf{k}-\mathbf{q}_2)$
\begin{eqnarray}\label{vertex8}
  \frac{1}{V}\sum_{\mathbf{q}_3}\left[ f(\mathbf{q}_2-\mathbf{q}_3) - f(-\mathbf{q}_3)  \right]
     e^{-i\mathbf{q}_3\cdot\mathbf{R}_l}
     \sum_{n\neq l} \sum_{k\neq l,k\neq n}
     \left[e^{-i(\mathbf{q}_2-\mathbf{q}_3)\cdot\mathbf{R}_{n}} -
       \overline{e^{-i(\mathbf{q}_2-\mathbf{q}_3)\cdot\mathbf{R}_{n}}}\right]
  \left[e^{-i\left(\mathbf{k}-\mathbf{q}_2\right)\cdot\mathbf{R}_k} -
    \overline{e^{-i\left(\mathbf{k}-\mathbf{q}_2\right)\cdot\mathbf{R}_k}}\right].
  \nonumber \\ 
\end{eqnarray}

We conclude that self-energy for the vertex function, $\Sigma^\mathcal{V}$ can be expressed in terms of the
following three-particle density,
\begin{eqnarray}\label{Anl3}
  && n_{l3}(\mathbf{q}_4,\mathbf{q}_2-\mathbf{q}_4,\mathbf{k}-\mathbf{q}_2) =
  e^{-i\mathbf{q}_4\cdot\mathbf{R}_l}
  \sum_{n\neq l} \sum_{k\neq l,k\neq n}
  \left[e^{-i(\mathbf{q}_2-\mathbf{q}_4)\cdot\mathbf{R}_{n}} -
    \overline{e^{-i(\mathbf{q}_2-\mathbf{q}_4)\cdot\mathbf{R}_{n}}}\right]
  \left[e^{-i\left(\mathbf{k}-\mathbf{q}_2\right)\cdot\mathbf{R}_k} -
    \overline{e^{-i\left(\mathbf{k}-\mathbf{q}_2\right)\cdot\mathbf{R}_k}}\right].
  \nonumber \\
\end{eqnarray}
Density $n_{l3}$ is a product of density of site $l$ and two fluctuations of collective densities
of all other particles.

\end{widetext}

\end{document}